\documentclass[3p,times,procedia]{elsarticle}

\usepackage{nupha_ecrc}
\usepackage{booktabs}
\usepackage{hyperref}
\usepackage{amsmath}

\newcommand{\pp}{\ensuremath{\rm{pp}} }

\newcommand{\PbPb}{\ensuremath{\text{Pb-Pb}} }
\newcommand{\snn}{\ensuremath{\sqrt{s_{\rm NN}}} }
\newcommand{\RAA}{\ensuremath{R_{\rm AA}} }

\newcommand{\TCF}{\ensuremath{T_{\rm CF}} }
\newcommand{\Ncoll}{\ensuremath{N_{\rm coll}} }
\newcommand{\der}{\ensuremath{ {\rm d}} }

\newcommand{\jpsi}{\ensuremath{\rm J  /\psi } }
\newcommand{\psiP}{\ensuremath{\rm \psi(2S)} }
\newcommand{\ccBar}{\ensuremath{\rm{c}\bar{\rm{c}}} }

\newcommand{\pT}{\ensuremath{p_{\rm T}} }

\newcommand{\mT}{\ensuremath{m_{\rm T}} }
\newcommand{\bT}{\ensuremath{\beta^{\rm s}_{\rm T}}}
\newcommand{\muB}{\ensuremath{\mu_{\rm B}} }

\volume{00}

\firstpage{1}

\journalname{Nuclear Physics A}

\jid{nupha}

\jnltitlelogo{Nuclear Physics A}

\usepackage{amssymb}

\usepackage[figuresright]{rotating}

\begin{document}

\begin{frontmatter}



\dochead{XXVIIth International Conference on Ultrarelativistic Nucleus-Nucleus Collisions\\ (Quark Matter 2018)}

\title{Testing charm quark thermalisation within the \\Statistical Hadronisation Model}


\author[mue]{A.~Andronic}
\author[gsi]{P.~Braun-Munzinger}
\author[hd]{M.~K.~K\"{o}hler}
\author[hd]{J.~Stachel}
\runauth{A.~Andronic et al.}

\address[mue]{Westf\"{a}lische Wilhelms-Universit\"{a}t M\"{u}nster, Institut f\"{u}r Kernphysik, M\"{u}nster, Germany}
\address[gsi]{Research Division and ExtreMe Matter Institute EMMI, GSI Helmholtzzentrum f\"{u}r \\Schwerionenforschung GmbH, Darmstadt, Germany}
\address[hd]{Physikalisches Institut, Ruprecht-Karls-Universit\"{a}t Heidelberg, Heidelberg, Germany}

\begin{abstract}
A wealth of data on charmonium production in \PbPb collisions from the LHC experiments has provided strong evidence for (re-)generation as a dominant production mechanism at low transverse momentum. We present an important extension of the statistical hadronisation model to describe \jpsi transverse momentum distributions based on input parameters from hydrodynamical simulations. Comparison to the data allows the testing of the degree of thermalisation of charm quarks in the quark-gluon plasma. To this end we will report analyses of the \jpsi transverse momentum spectra in \PbPb collisions at $\snn = 2.76$ and $5.02$~TeV.
\end{abstract}


\end{frontmatter}


%
%
\section{Introduction}
\label{sec:Intro}
Charmonium production has proven to be an intriguing probe for the hot and dense system produced in ultrarelativistic heavy-ion collisions. The suppression~\cite{MatsuiSatz:1986} and (re-)combination mechanism~\cite{BraunMunzinger:2000px,Thews:2000rj} reflect the underlying dynamics of charm and anti-charm quarks in the quark-gluon plasma (QGP) and at the phase boundary. In central heavy-ion collisions, the suppression of charmonium production compared to the vacuum expectation showed a significant weakening for increasing collision energies~\cite{PhysRevLett.109.072301,Abelev:2013ila}. This could be explained by the increasing importance of the recombination mechanism due to the increased charm cross section~\cite{Andronic:2006ky,ZhaoRapp:2011,Ferreiro:2012rq}. The large amount of recombined charmonium should also be reflected in the charmonium kinematics and in particular in the transverse momentum distribution~\cite{Andronic:2006ky,ZhaoRapp:2011}. \\
In this contribution, we report on charmonium production in \PbPb collisions at LHC energies calculated within the framework of the statistical hadronisation model (SHM). The centrality, rapidity and transverse momentum dependence will be compared with available data from LHC. With the implementation of the transverse momentum spectra in the SHM, the degree of thermalisation of charm quarks in the QGP at the critical temperature is tested.
%
%
\section{Heavy quarks in the statistical hadronisation model}
\label{sec:shm}
The SHM assumes that heavy quarks are produced in initial hard scatterings and that thermal production is negligible at current energies~\cite{Andronic:2006ky}. All produced heavy-flavour quarks survive in and thermalise within the QGP. Above the chemical freeze-out temperature, $\TCF$, all hadrons are fully screened and no colour-less bound states exist in the fireball \mbox{volume $V$}. Charmonium, together with all the other hadrons, is formed at the phase boundary, $T = \TCF$. Hadron yields can be described within the grand-canonical ensemble at $\TCF = 156.5$~MeV and vanishing baryon chemical potential $\muB$, see e.g.~\cite{Andronic:2017} for a recent review. \\
The number of produced $\ccBar$ pairs $N_{\ccBar}$ in a collision is linked to the statistical ensemble yields via
\begin{equation}
  N_{\ccBar} = \frac{1}{2} g_c V \left\{ \sum_i \left( n^{\rm th}_{D_i} + n^{\rm th}_{\Lambda_i} + \cdots \right) \right\} + g_c^2 V\left\{  \sum_i \left( n^{\rm th}_{\psi_i} + n^{\rm th}_{\chi_i} + \cdots \right) \right\} ,
  \label{eq:balance}
\end{equation}
where $g_c$ is a charm quark fugacity and $n^{\rm th}_X$ are the particle and anti-particle densities in the fireball \mbox{volume $V$}. The amount of produced $\ccBar$ pairs is given by the total charm cross section in a nucleus-nucleus collision $\der \sigma^{\rm AA}_{\ccBar} / \der y$. It should be emphasised that $\der \sigma^{\rm AA}_{\ccBar} / \der y$ is the only additional input parameter needed for the calculations of charmonium yields in the SHM. However, there are no measurements for $\der \sigma^{\rm AA}_{\ccBar} / \der y$ so far, so the quantity is estimated from the corresponding charm cross sections in \pp collisions, $\der \sigma^{\pp}_{\ccBar} / \der y$, in the corresponding rapidity region~\cite{LHCb:5TeV,LHCb:7TeV,LHCb:13TeV,ALICEccBar:2017}, where the shape of $\der \sigma^{\pp}_{\ccBar} / \der y$ is estimated by FONLL~\cite{Cacciari:2012ny,Cacciari:2015fta} calculations. Scaling is done via the nuclear overlap function~\cite{Andronic:2006ky}. The impact of shadowing $S(y)$ is estimated from the nuclear modification factor in p-Pb collisions by using the geometrical relation $S(y) = R_{\rm pPb}(y) \times R_{\rm pPb} (-y)$, where $R_{\rm pPb}(y)$ is taken from \jpsi and $D$ meson measurements~\cite{Adam:2015iga,Aaij:2017gcy} and a shape interpolation, if necessary, is done using model calculations in~\cite{Aaij:2017gcy}.
In the balance equation (\ref{eq:balance}) the first term relates to open charm hadrons and the second term to charmonia, hence the linear or squared appearance of the fugacity. Higher order charmed particles can be neglected~\cite{Andronic:2006ky}. A canonical correction factor $I_1(g_c n^{\rm th}_{\rm oc} V) / I_0(g_c n^{\rm th}_{\rm oc} V)$ is applied to the open charm term, where $I_n$ are mo\-dified Bessel functions. This correction gains importance towards peripheral collisions when the number of \ccBar pairs is small, $N_{\ccBar} \lesssim 1$. Through equation (\ref{eq:balance}) the value for the fugacity is fully determined.\\
It has to be taken into account, that nucleons from the surface of the colliding nuclei can be assumed not to contribute to the fireball since they undergo one or zero nucleon-nucleon scatterings. Nucleons are therefore separated into a ``core'' part, which contributes to the thermal charmonium production in fireball, $N^{\rm core}_{\jpsi} = g_c^2 n^{\rm th}_{\jpsi}V$, and a ``corona'' part, $N^{\rm corona}_{\jpsi} = \Ncoll ^{\rm corona} \times \sigma^{\pp}_{\jpsi}/\sigma^{\pp}_{\rm inel}$, which is treated like individual \pp collisions. The total amount of \jpsi is then given by $N_{\jpsi} = N^{\rm core}_{\jpsi} + N^{\rm corona}_{\jpsi}$.\\
While $\TCF$ and $\muB$ are rapidity independent, the rapidity dependence of the fireball volume is estimated by $V(y) = \der N_{\rm ch} / \der y \hspace{0.1 cm} / \hspace{0.1 cm} n^{\rm th}_{\rm ch}$, where $\der N_{\rm ch} / \der y$ is the charged particle rapidity distribution at the corresponding collision energy and centrality~\cite{ALICE:dNchdy2760,ALICE:dNchdy5020} and $n^{\rm th}_{\rm ch}$ is the charged particle density from the SHM. \\
The underlying assumption, that thermalised charm quarks form charmonia at the chemical freeze-out temperature, can be extended to calculate transverse momentum spectra. Then, charm quarks follow the collective expansion of the fireball, which is known to be modelled well by viscous hydrodynamical simulations for the light flavour sector~\cite{Song:2010aq,Gale:2013}. MUSIC(3+1)D~\cite{Schenke:2010nt} hydrodynamical simulations are used with QCD-based parameters~\cite{Dubla:2018czx} and IP-Glasma~\cite{Schenke:2012wb} as initial conditions to model the freeze-out hyper surface at $T=\TCF$. The results of the hydrodynamical simulations are used to constrain the blast-wave function~\cite{Schnedermann:1993ws}
\begin{equation}
  \frac{\der N}{\pT \der \pT} \propto \int_0^R \der r \hspace{0.1 cm} r \mT I_0 \left( \frac{\pT \sinh \rho}{T} \right) K_1 \left( \frac{\mT \cosh \rho}{T}\right),
\end{equation}
where $\rho = \tanh^{-1} \left\{ \bT (r/R)^n\right\}$ with the transverse velocity at the freeze-out hyper surface \bT, the radial velocity profile $n$, which is found to be close to unity, and the modified Bessel functions $I_0$ and $K_1$. We have also used an analogous blast wave formula differential in $y$ and $\pT$~\cite{Andronic_to_be_published}, see also a different version in section $25.2.2$ of~\cite{Florkoski_book}, but the differences to what is shown here are small. The resulting blast-wave function is used to model the shape of the thermal part of the transverse momentum spectrum which is normalised to the core fraction given by the SHM. The shape of the corona part is modelled by \jpsi measurements in \pp collisions at forward rapidity~\cite{Abelev:2012kr,Acharya:2017hjh} and by an interpolation procedure at mid-rapidity~\cite{Bossu:2011qe}.
%
%
\section{Results}
\label{sec:results}
\begin{figure}[t]
  \centering
  \includegraphics[scale = 0.37]{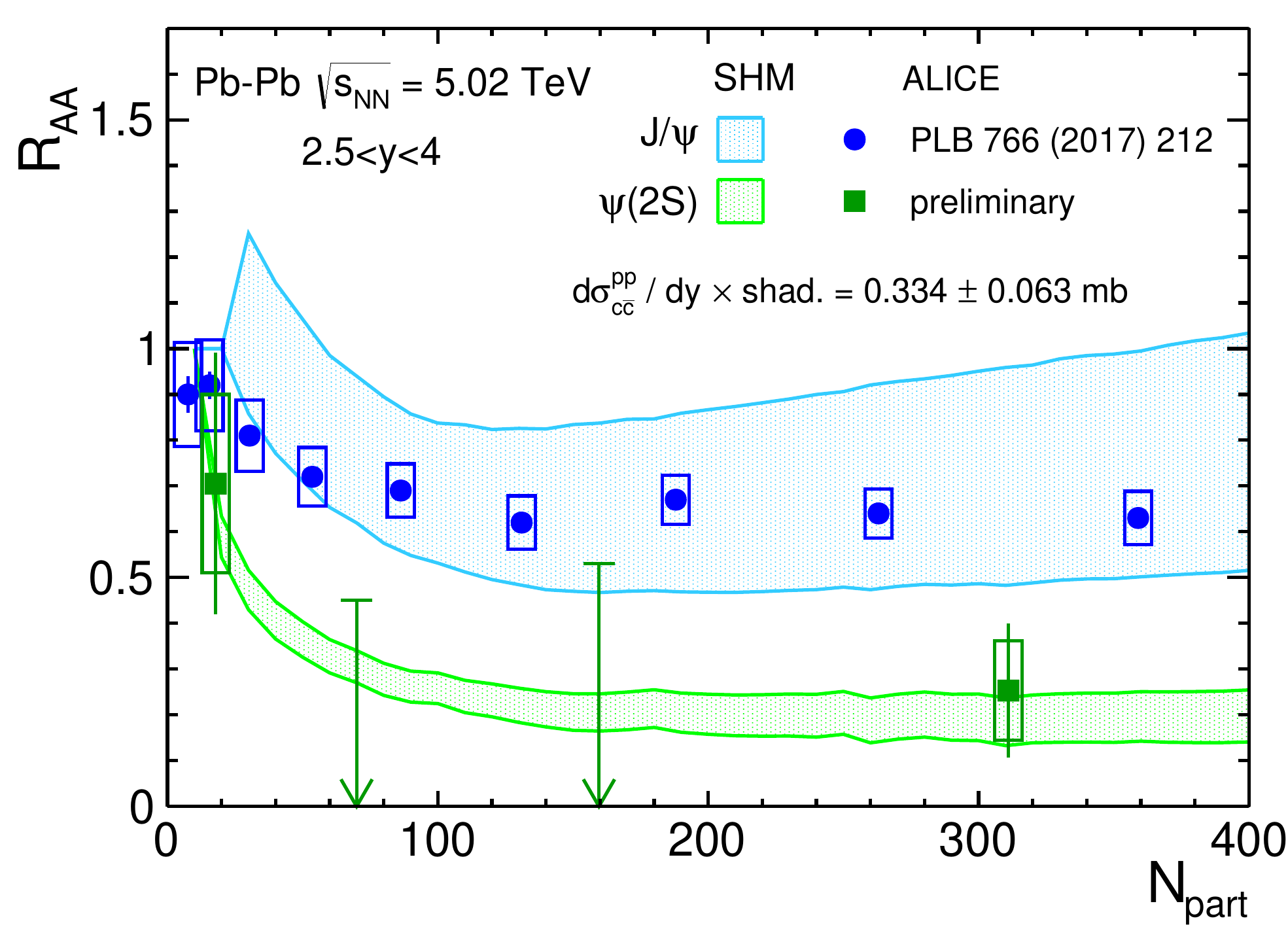}
  \includegraphics[scale = 0.37]{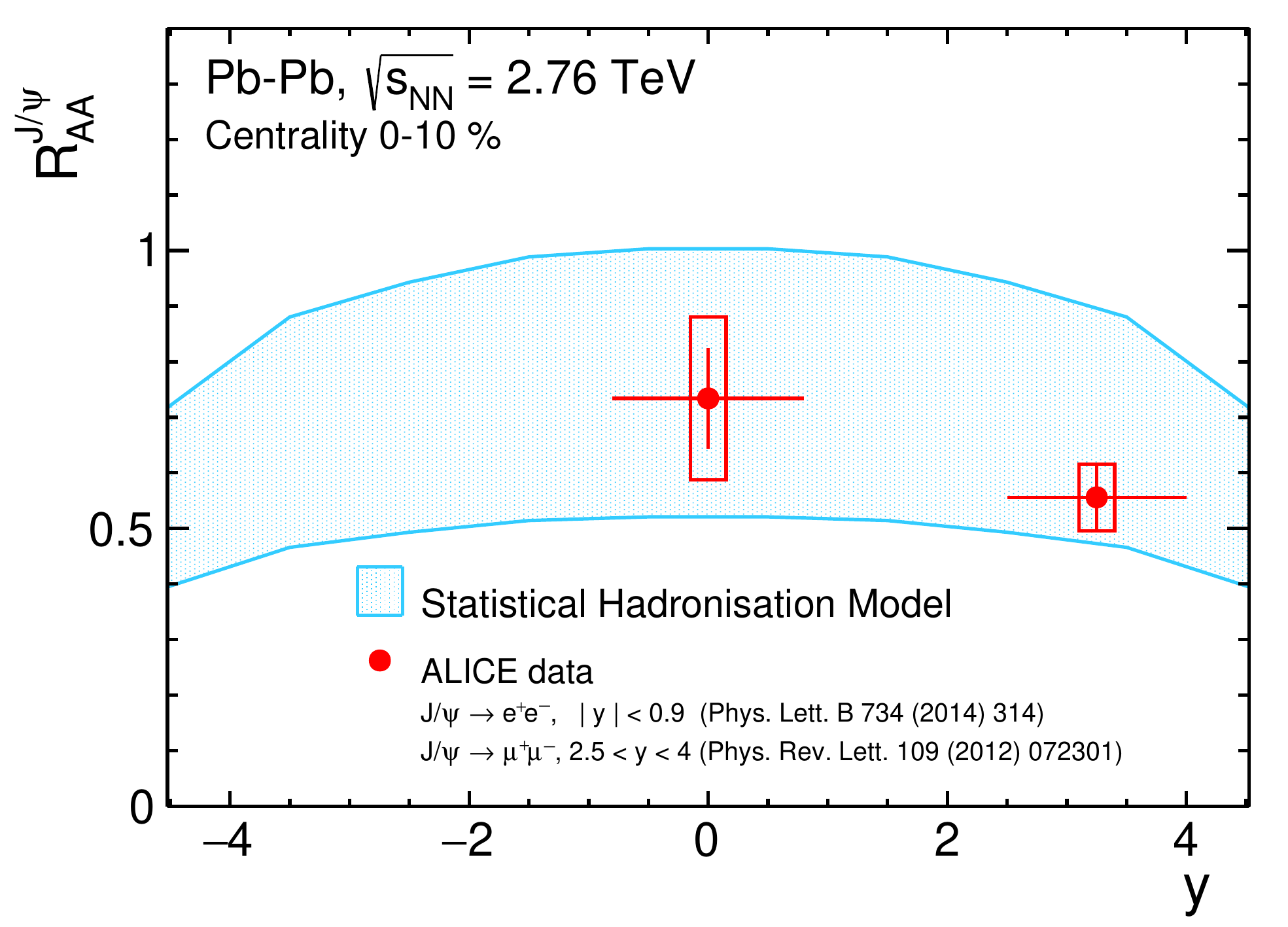}
  \caption{Results on the nuclear modification factor \RAA from the SHM are compared to LHC data. In the left panel results are shown for \jpsi and \psiP as a function of the centrality at a collision energy of $\snn = 5.02$~TeV and compared to data~\cite{Adam:2016rdg,ALICE:Tarhini}. In the right panel, the model is compared to data~\cite{PhysRevLett.109.072301,Abelev:2013ila} as a function of rapidity in the most central collisions \mbox{($0-10$~\%)} at a collision energy of $\snn = 2.76$~TeV.}\label{fig:RAA_cent_rap}
\end{figure}

\begin{figure}[t]
  \centering
  \includegraphics[scale = 0.37]{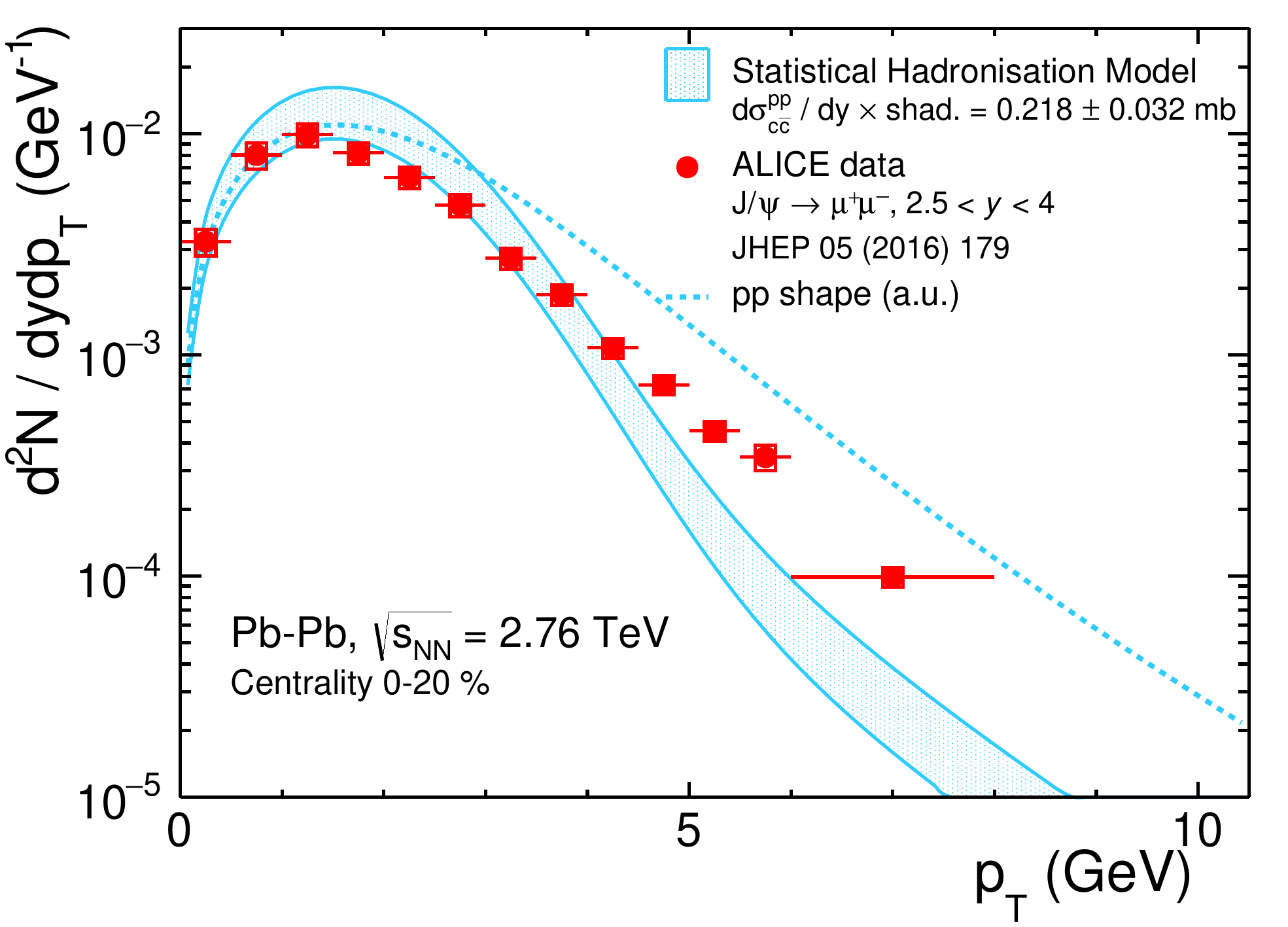}
  \includegraphics[scale = 0.37]{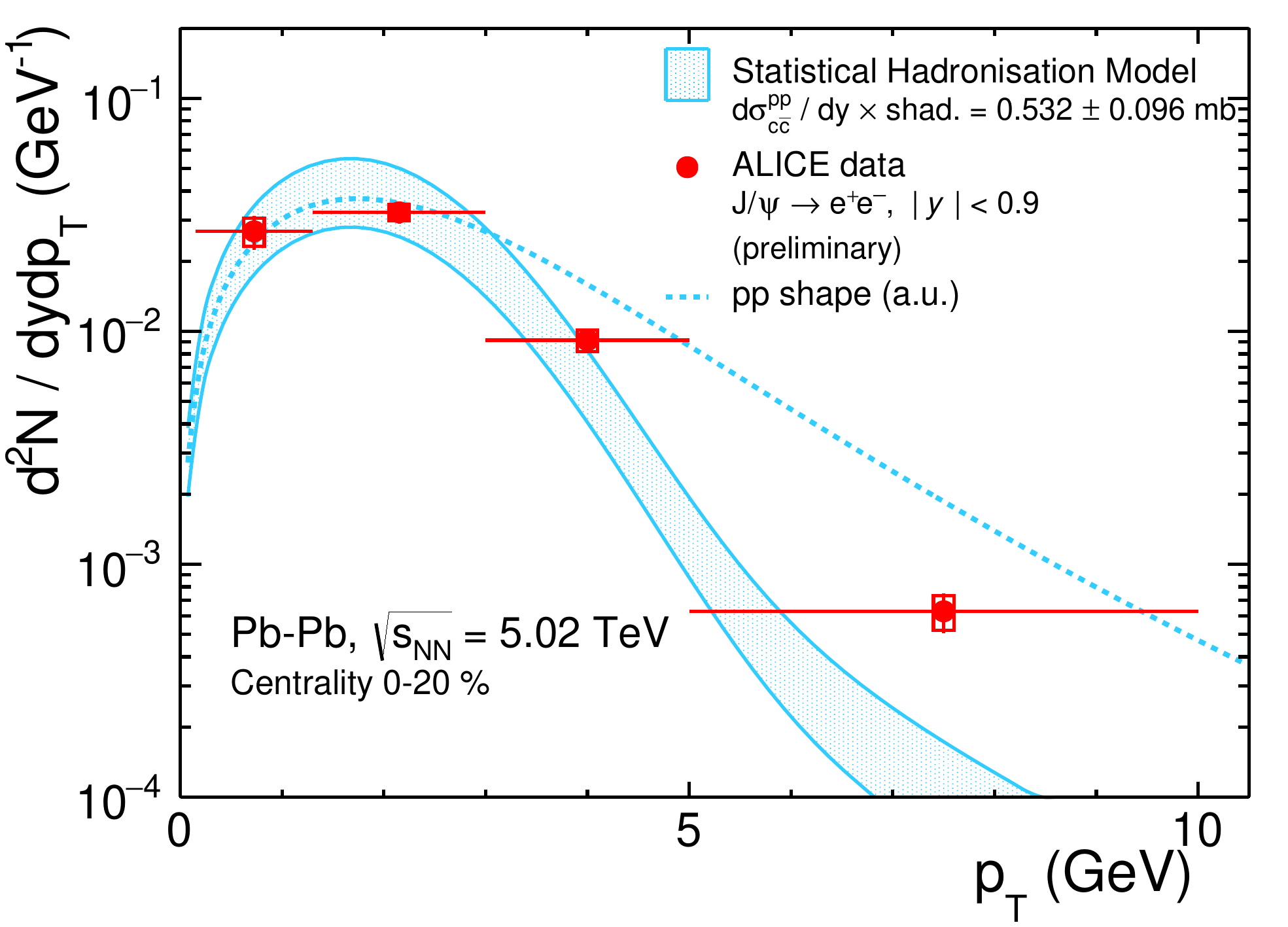}
  \caption{Results from the SHM for transverse momentum spectra at forward rapidity at $\snn = 2.76$~TeV (left panel) and at mid-rapidity at $\snn = 5.02$~TeV (right panel) are shown for the most central collisions \mbox{($0-20$~\%)} as bands and compared to available data~\cite{ALICE:JPsiForw2760,ALICE:Dennis}. The arbitrarily normalised \pT spectrum from \pp collisions is added to emphasise the difference to the full shape obtained by the procedure described in the text. The shape of the \pT spectra in pp collisions is extracted by a fit to available data~\cite{Abelev:2012kr} in the case of forward rapidity and by an interpolation procedure in case of mid-rapidity~\cite{Bossu:2011qe}.}\label{fig:pt_spec}
\end{figure}

In the left panel of Fig.~\ref{fig:RAA_cent_rap}, the nuclear modification factors \RAA of the charmonium states \jpsi and \psiP are shown for forward rapidity as a function of the centrality. The result of the SHM is shown as a band with a width determined mostly by the shadowing uncertainty. The SHM calculations for both charmonium states are compared to data~\cite{Adam:2016rdg,ALICE:Tarhini} at the corresponding collision energy and show very good agreement.\\
The \RAA of \jpsi as a function of rapidity for the most central collisions ($0-10$~\%) at $\snn = 2.76$~TeV is shown in the right panel of Fig.~\ref{fig:RAA_cent_rap}. The model describes the data~\cite{PhysRevLett.109.072301,Abelev:2013ila} very well; it should be emphasised that the rapidity dependence of \jpsi calculated within the SHM is given by the shape of the charm cross section and follows naturally the trend of the data, i.e. is decreasing towards larger rapidities, due to the dilution of charm quarks towards larger rapidities. This is in contrast to screening dominated models.\\
In Fig.~\ref{fig:pt_spec}, the transverse momentum spectrum at forward-rapidity  at \mbox{$\snn = 2.76$~TeV} (left panel) and mid-rapidity at \mbox{$\snn = 5.02$~TeV} (right panel) for the centrality \mbox{$0-20$~\%} from the SHM is compared to ALICE data~\cite{ALICE:JPsiForw2760,ALICE:Dennis}. The agreement at low \pT is very good, where for \mbox{$\pT \gtrsim 5$~GeV} an overshoot of the data compared to the model can be seen.\\ 
The overshoot at high \pT indicates, that another production mechanism is gaining importance towards high \pT, reminiscent to the behaviour observed for open charmed mesons~\cite{ALICE:Dmeson2018} and charged particles~\cite{ATLAS:Jpsi2018}.
%
%
\section{Summary and conclusions}
\label{sec:summary}
We presented results on charmonium production within the framework of the statistical hadronisation model as a function of centrality, rapidity and transverse momentum assuming full thermalisation in the QGP, constrained with state-of-the-art hydrodynamic modelling. The model showed a very good agreement with available data as a function of centrality, rapidity and in particular at low transverse momentum. The good agreement strongly supports the picture, that charmonia at low and moderate transverse momentum are formed at the phase boundary from deconfined charm quarks flowing with the quark-gluon plasma. To date this is the most convincing demonstration that the medium formed consists of deconfined quarks. The discrepancy at higher \pT suggests that an additional production mechanism is gaining importance.

\section*{Acknowledgment}
We thank A.~Dubla, K.~Reygers and C.~Shen for fruitful discussions. This work is part of and supported by the DFG Collaborative Research Centre  ``SFB 1225 (ISOQUANT)".

%
%

\end{document}